%% file: rsc-articletemplate.tex
\documentclass[twoside,twocolumn,9pt]{article}
\usepackage{extsizes}
\usepackage[super,sort&compress,comma]{natbib} 
\usepackage[version=3]{mhchem}
\usepackage[left=1.5cm, right=1.5cm, top=1.785cm, bottom=2.0cm]{geometry}
\usepackage{balance}
\usepackage{mathptmx}
\usepackage{sectsty}
\usepackage{graphicx} 
\usepackage{lastpage}
\usepackage[format=plain,justification=justified,singlelinecheck=false,font={stretch=1.125,small,sf},labelfont=bf,labelsep=space]{caption}
\usepackage{float}
\usepackage{fancyhdr}
\usepackage{fnpos}
\usepackage[english]{babel}
\addto{\captionsenglish}{%
  
}
\usepackage{array}
\usepackage{droidsans}
\usepackage{charter}
\usepackage[T1]{fontenc}
\usepackage[usenames,dvipsnames]{xcolor}
\usepackage{setspace}
\usepackage[compact]{titlesec}
\usepackage{hyperref}
 \usepackage{textcomp}

\usepackage{epstopdf}

\definecolor{cream}{RGB}{222,217,201}

\begin{document}

\pagestyle{fancy}
\thispagestyle{plain}
\fancypagestyle{plain}{
\renewcommand{\headrulewidth}{0pt}
}

\makeFNbottom
\makeatletter
\renewcommand\LARGE{\@setfontsize\LARGE{15pt}{17}}
\renewcommand\Large{\@setfontsize\Large{12pt}{14}}
\renewcommand\large{\@setfontsize\large{10pt}{12}}
\renewcommand\footnotesize{\@setfontsize\footnotesize{7pt}{10}}
\makeatother

\renewcommand{\thefootnote}{\fnsymbol{footnote}}
\renewcommand\footnoterule{\vspace*{1pt}%
\color{cream}\hrule width 3.5in height 0.4pt \color{black}\vspace*{5pt}} 
\setcounter{secnumdepth}{5}

\makeatletter 
\renewcommand\@biblabel[1]{#1}            
\renewcommand\@makefntext[1]%
{\noindent\makebox[0pt][r]{\@thefnmark\,}#1}
\makeatother 
\renewcommand{\figurename}{\small{Fig.}~}
\sectionfont{\sffamily\Large}
\subsectionfont{\normalsize}
\subsubsectionfont{\bf}
\setstretch{1.125} 
\setlength{\skip\footins}{0.8cm}
\setlength{\footnotesep}{0.25cm}
\setlength{\jot}{10pt}
\titlespacing*{\section}{0pt}{4pt}{4pt}
\titlespacing*{\subsection}{0pt}{15pt}{1pt}

\fancyfoot{}
\fancyfoot[LO,RE]{\vspace{-7.1pt}\includegraphics[height=9pt]{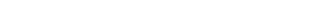}}
\fancyfoot[CO]{\vspace{-7.1pt}\hspace{13.2cm}\includegraphics{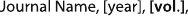}}
\fancyfoot[CE]{\vspace{-7.2pt}\hspace{-14.2cm}\includegraphics{head_foot/RF}}
\fancyfoot[RO]{\footnotesize{\sffamily{1--\pageref{LastPage} ~\textbar  \hspace{2pt}\thepage}}}
\fancyfoot[LE]{\footnotesize{\sffamily{\thepage~\textbar\hspace{3.45cm} 1--\pageref{LastPage}}}}
\fancyhead{}
\renewcommand{\headrulewidth}{0pt} 
\renewcommand{\footrulewidth}{0pt}
\setlength{\arrayrulewidth}{1pt}
\setlength{\columnsep}{6.5mm}
\setlength\bibsep{1pt}

\makeatletter 
\newlength{\figrulesep} 
\setlength{\figrulesep}{0.5\textfloatsep} 

\newcommand{\topfigrule}{\vspace*{-1pt}%
\noindent{\color{cream}\rule[-\figrulesep]{\columnwidth}{1.5pt}} }

\newcommand{\botfigrule}{\vspace*{-2pt}%
\noindent{\color{cream}\rule[\figrulesep]{\columnwidth}{1.5pt}} }

\newcommand{\dblfigrule}{\vspace*{-1pt}%
\noindent{\color{cream}\rule[-\figrulesep]{\textwidth}{1.5pt}} }

\makeatother

\twocolumn[
  \begin{flushleft}
{\includegraphics[height=30pt]{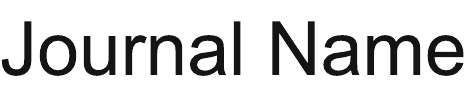}\\[1ex]
\includegraphics[width=18.5cm]{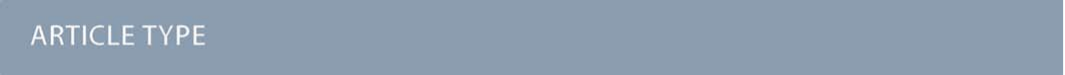}}\par
\vspace{1em}
\sffamily
\begin{tabular}{m{4.5cm} p{13.5cm} }

\includegraphics{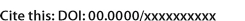} & \noindent\LARGE{\textbf{SABRE Hyperpolarization of Unmodified Amino Acids in Partially Aqueous Media: L-[1-\textsuperscript{13}C]-Valine as a Model System.}} \\
\vspace{0.3cm} & \vspace{0.3cm} \\

 & \noindent\large{Oksana A. Bondar$^{\ast}$\textit{$^{a}$}} and Thomas B. R. Robertson$^{\textit{$^{a}$}}$ \\

\includegraphics{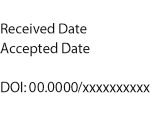} & \noindent\normalsize{Hyperpolarization techniques enhance the sensitivity of nuclear magnetic resonance (NMR) and magnetic resonance imaging (MRI), enabling detection of low-concentration metabolites and dynamic processes. Signal Amplification by Reversible Exchange (SABRE) transfers spin polarization from parahydrogen without permanent chemical modification of the substrate.

Here, we demonstrate SABRE-mediated \textsuperscript{13}C hyperpolarization of unmodified amino acids using L-[1-\textsuperscript{13}C]-valine as the primary model system and provide preliminary evidence of applicability to glycine. Signal enhancements exceeding 60-fold were observed on a 1.1 T benchtop NMR spectrometer. The dependence of polarization on parahydrogen bubbling time, solvent composition, and substrate-to-catalyst ratio was investigated.} \\

\end{tabular}

 \end{flushleft} \vspace{0.6cm}
  ]

\renewcommand*\rmdefault{bch}\normalfont\upshape
\rmfamily
\section*{}
\vspace{-1cm}


\footnotetext{\textit{$^{a}$~School of Chemistry, Highfield Campus, Southampton, SO17 1BJ United Kingdom; E-mail: oksana14.bondar@gmail.com}}

\footnotetext{\dag~Electronic Supplementary Information (ESI) available: [details of any supplementary information available should be included here]. See DOI: 00.0000/00000000.}



\section{Introduction}

Nuclear magnetic resonance (NMR) spectroscopy is a powerful analytical technique widely used for molecular structure determination, reaction monitoring, and metabolic studies. However, its inherently low sensitivity, arising from the small Boltzmann population difference between nuclear spin states at thermal equilibrium, remains a major limitation\cite{ardenkjaer2015facing}.To address this challenge, a variety of hyperpolarization methods have been developed to increase nuclear spin polarization by several orders of magnitude above thermal equilibrium, enabling applications in metabolic imaging and real-time studies of biochemical processes\cite{ardenkjaer2015facing,kurhanewicz2019hyperpolarized,cavallari2020vitro}.

Among these approaches, Signal Amplification by Reversible Exchange (SABRE), originally reported by Adams et al.\cite{adams2009reversible}, has emerged as a powerful hyperpolarization technique because it enables the transfer of spin polarization from parahydrogen to target molecules without permanent chemical modification\cite{duckett2012application,barskiy2019sabre}.

Following the original demonstration of SABRE, subsequent work expanded the methodology beyond pyridine-based systems to a broader range of substrates and heteronuclei\cite{duckett2012application,barskiy2019sabre}. In particular, Glöggler and co-workers demonstrated \textsuperscript{13}C SABRE hyperpolarization of diverse molecular substrates and provided an early demonstration of extending SABRE beyond the original pyridine-based systems\cite{gloggler2011hydrogen}.

SABRE relies on the reversible binding of both parahydrogen and a substrate to a transition-metal catalyst, typically an iridium complex, allowing polarization transfer through scalar (J) coupling interactions within the transient catalyst–substrate complex\cite{adams2009reversible,duckett2012application}.

Compared with dissolution DNP, SABRE does not require cryogenic temperatures, microwave irradiation, or a dissolution step. Instead, polarization is generated through reversible substrate–catalyst exchange with parahydrogen under ambient-temperature conditions, enabling rapid, repeatable polarization experiments using conventional NMR instrumentation\cite{gloggler2015hyperpolarization}.

Despite these advantages, the application of SABRE to biologically relevant molecules remains challenging. Unmodified amino acids are attractive targets for hyperpolarized NMR because they participate directly in central metabolic pathways and can serve as endogenous biomarkers of physiological and pathological processes. Their zwitterionic nature leads to coordination behavior that differs from the nitrogen-containing substrates traditionally used in SABRE, while the presence of water can adversely affect catalyst stability and polarization transfer efficiency\cite{kaltschnee2019hyperpolarization,gater2025perfluorosolvents,min2025waterrecyclablecatalysts}.

\begin{figure}
    \centering
    \includegraphics[width=0.45\linewidth]{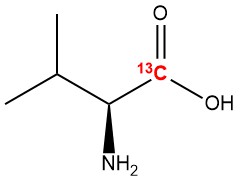}
    \caption{Graphical picture of L-[1-\textsuperscript{13}C]-Valine}
    \label{pic_Val}
\end{figure}

To address these challenges, we present a SABRE protocol employing mixed organic--aqueous solvent systems to balance catalyst solubility and polarization transfer efficiency for the direct \textsuperscript{13}C hyperpolarization of unmodified amino acids, using L-[1-\textsuperscript{13}C]-valine (Fig.~\ref{pic_Val}) as a model system. Among the branched-chain amino acids, L-valine is of particular interest because it plays important roles in energy metabolism and has been investigated as a metabolic biomarker in cancer and neurological disorders\cite{chaumeil2024new,papaneophytou2024warburg,put2023detection}. Recent studies have demonstrated that unmodified $\alpha$-amino acids can participate in parahydrogen-based hyperpolarization experiments and interact with Ir-based hyperpolarization catalysts, highlighting the potential of SABRE for amino acid detection and characterization\cite{sellies2021parahydrogen}.

While valine has previously been hyperpolarized using dissolution dynamic nuclear polarization (DNP)\cite{soon2013hyperpolarization}, SABRE enables rapid polarization generation without the need for cryogenic temperatures, microwave irradiation, or a dissolution step, although the achievable polarization levels depend strongly on the substrate and experimental conditions. Accordingly, this work investigates the feasibility of direct SABRE-mediated \textsuperscript{13}C hyperpolarization of unmodified L-[1-\textsuperscript{13}C]-valine in a partially aqueous solvent system. Particular attention is given to the influence of bubbling time, solvent composition, and substrate-to-catalyst ratio on the observed signal enhancement, providing practical insight into the experimental factors governing SABRE hyperpolarization of unmodified amino acids. The influence of solvent composition observed in the present study is consistent with previous reports demonstrating that solvent choice strongly affects catalyst performance and hyperpolarization efficiency in parahydrogen-based hyperpolarization methods\cite{bondar2022effect}.

This study presents a systematic investigation of direct SABRE-mediated \textsuperscript{13}C hyperpolarization of an unmodified amino acid in a partially aqueous solvent system, with particular emphasis on the influence of experimental parameters governing polarization efficiency.

\section{Results and Discussion}

\subsection{Hyperpolarization of L‑[1‑\textsuperscript{13}C]‑Valine via SABRE and Coordination Heterogeneity}

We demonstrate the SABRE hyperpolarization of unmodified L-[1-\textsuperscript{13}C]-Valine in a partially aqueous solvent system with signal detection at 1.1 T using a benchtop NMR spectrometer. Samples containing the amino acid and the iridium precatalyst were exposed to 92 $\%$ enriched parahydrogen for 20 s in a $\mu$-metal shielded environment (B\textsubscript{0} $\approx$ 9 mG) at room temperature.

\begin{figure} [ht]
\includegraphics[width=0.45\textwidth]{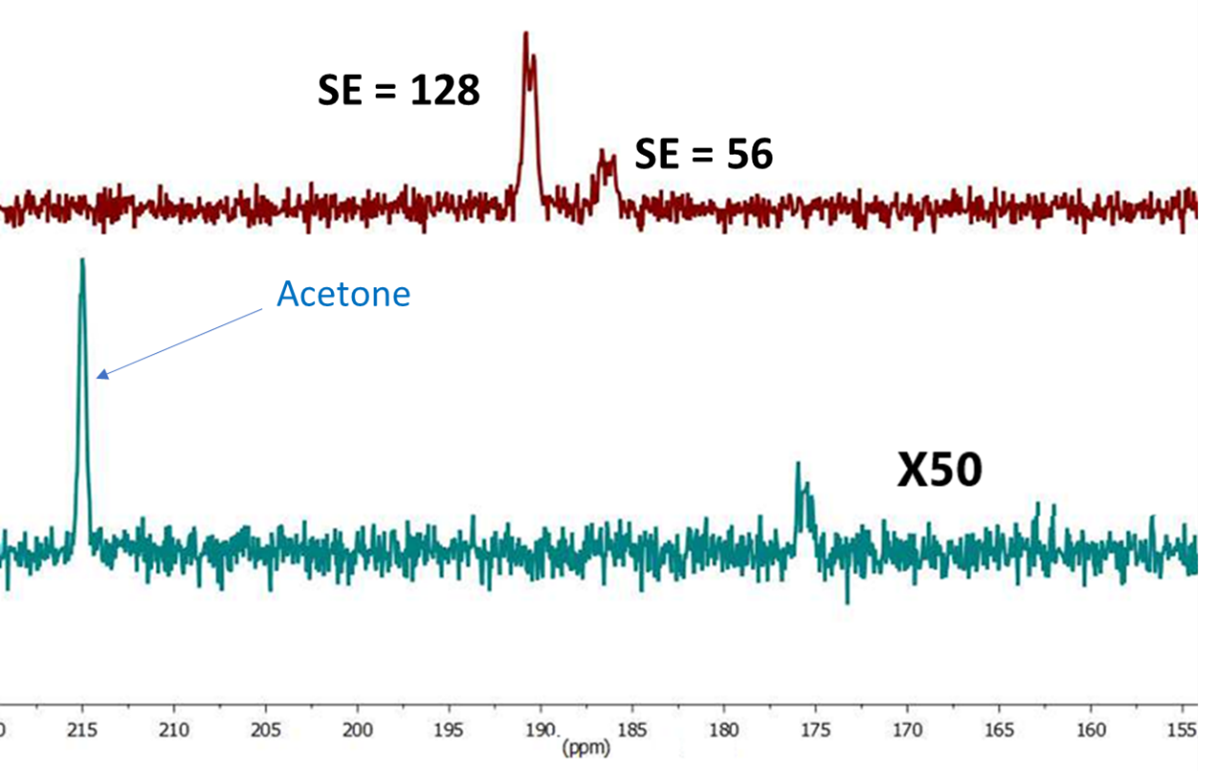}
\caption{SABRE hyperpolarization of unmodified L-[1-\textsuperscript{13}C]-Valine. Comparison of \textsuperscript{13}C NMR spectra acquired at 1.1 T: bottom, thermal spectrum (1024 scans, x50 vertical scaling); top, SABRE-hyperpolarized spectrum (1 scan).The hyperpolarized spectrum shows clear enhancement of the \textsuperscript{13}C resonances corresponding to catalyst-associated valine species relative to the thermal spectrum (SE $\approx$ 128), with a secondary resonance (SE $\approx$ 56). Acetone peak marked for reference.}
\label{HP_valine}
\end{figure}

Figure~\ref{HP_valine} compares the thermal \textsuperscript{13}C spectrum obtained after signal averaging (1024 scans) with the hyperpolarized spectrum acquired in a single scan. Hyperpolarization results in two enhanced \textsuperscript{13}C resonances at approximately 190 and 186 ppm, whereas the thermal spectrum contains only the resonance of free L-[1-\textsuperscript{13}C]-valine at approximately 175 ppm.

The resonance at approximately 175 ppm is assigned to the carboxyl carbon of free L-[1-\textsuperscript{13}C]-valine. Because the \textsuperscript{13}C isotope label is located exclusively at the C1 carboxyl carbon, and no other \textsuperscript{13}C-enriched species were present in the sample, the observed hyperpolarized resonances are most reasonably attributed to species containing the labeled valine carbon. Their downfield shift relative to the free-valine resonance is consistent with a change in the electronic environment of the carboxyl carbon, which may arise from interaction with the iridium catalyst. Previous computational and experimental studies have shown that amino acids, including valine, can coordinate to Ir-IMes SABRE catalysts and form multiple catalyst-associated species\cite{bouma2023computational}. The observation of two hyperpolarized resonances is therefore qualitatively consistent with the presence of multiple catalyst-associated valine environments. However, the present data do not permit definitive structural assignment of these catalyst-associated species, and additional spectroscopic or structural studies would be required to determine their precise coordination environments.

The signal enhancements observed in this study are modest compared with the highest SABRE enhancements reported for optimized nitrogen-containing substrates, which can exceed several orders of magnitude under favorable conditions. Direct SABRE hyperpolarization of unmodified amino acids has proven considerably more difficult than for the nitrogen-containing heterocycles traditionally used in SABRE because amino acids interact differently with the iridium catalyst and must often be studied in the presence of water. Within this context, the observation of reproducible \textsuperscript{13}C signal enhancement from unmodified L-[1-\textsuperscript{13}C]-valine in a partially aqueous solvent system provides evidence that SABRE polarization transfer can be achieved in amino-acid-based systems without chemical derivatization.

\subsection{Effect of Parahydrogen Bubbling Time on Signal Enhancement}

The efficiency of SABRE hyperpolarization depends critically on dynamic exchange between parahydrogen, the iridium catalyst, and the substrate. To investigate how the duration of parahydrogen bubbling influences signal enhancement (SE), we measured the \textsuperscript{13}C SE of L-[1-\textsuperscript{13}C]-Valine at various bubbling times while keeping the [Val]/[Cat] ratio fixed.
Because no corresponding thermal resonances were observed at 190 and 186 ppm, apparent signal enhancements were estimated relative to the detectable thermal resonance of free L-[1-\textsuperscript{13}C]-valine observed near 175 ppm. The calculated values should therefore be regarded as approximate enhancement estimates rather than direct measurements of polarization gain for the individual hyperpolarized species.

\begin{figure} [ht]
\includegraphics[width=0.45\textwidth]{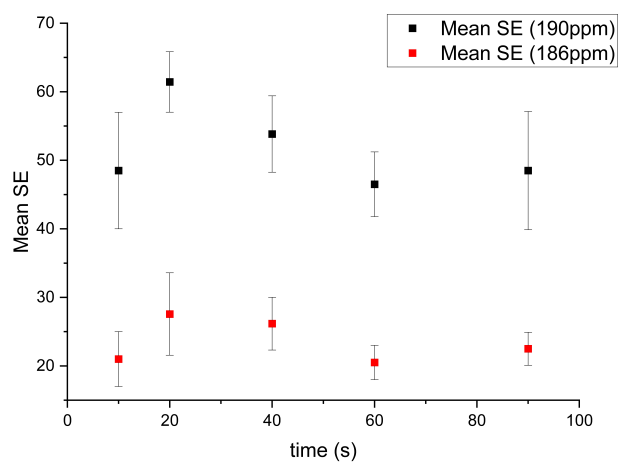}
\caption{Signal enhancement (SE) as a function of parahydrogen bubbling time for the two observed \textsuperscript{13}C resonances ($\sim$190 ppm and $\sim$186 ppm) of L-[1-\textsuperscript{13}C]-valine under [Val]/[Cat] = 9:1 in acetone:D\textsubscript{2}O (2:1 v/v). Error bars represent the standard deviation of repeated measurements. The number of replicates for each data point is provided in Table \ref{table_S1} (ESI).}
\label{SE_vs_btime}
\end{figure}

As shown in Figure\ref{SE_vs_btime}, signal enhancement increases with bubbling time up to approximately 20 seconds under the conditions studied ([Val]/[Cat] = 9:1), after which it decreases. While some variability is observed, this trend is consistent across repeated measurements.
A summary of the signal enhancement values, associated standard deviations, and number of replicates is provided in Table\ref{table1}. The corresponding individual measurements are reported in Tables \ref{table_S1} and \ref{table_S2} (ESI). 

\begin{table*} [ht]
\small
\centering
\caption{Summary of signal enhancement (SE) values as a function of parahydrogen bubbling time for the two observed \textsuperscript{13}C resonances ($\sim$190 ppm and $\sim$186 ppm) of L-[1-\textsuperscript{13}C]-Valine (Figure\ref{SE_vs_btime}).}
\label{table1}
    \begin{tabular}{c|c|c|c|c|c}
 \hline\
  Bubbling time & SE(bound Valine at 190 ppm) & Error SE at 190 ppm &  SE(bound Valine at 186 ppm) & Error SE at 186 ppm & n(repeats) \\  
 \hline\hline
 10	& 48.5	& 8.5	& 21	& 4	& 2 \\
\hline\
20	& 61.42857	& 4.40959	& 27.57143	& 6.00925 & 7 \\
\hline
40	& 53.83333	& 5.56776	& 26.16667	& 3.84419	& 6 \\ 
\hline
60 & 46.5	& 4.72582	& 20.5	& 2.5	& 4 \\
\hline
90 & 48.5	& 8.62168	& 22.5	& 2.4037	& 4 \\
\hline
\end{tabular}
\end{table*}

\begin{figure} [ht]
\includegraphics[width=0.5\textwidth]{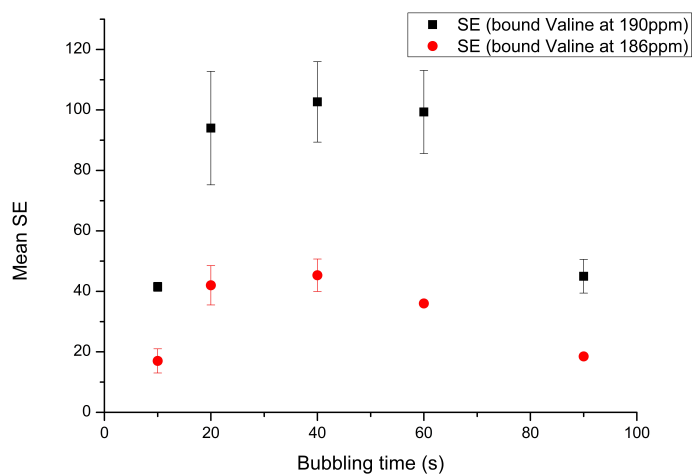}
\caption{Signal enhancement (SE) as a function of parahydrogen bubbling time (10–90 s) for a substrate-to-catalyst ratio of [Val]/[Cat]=6:1. Both \textsuperscript{13}C resonances of L-[1-\textsuperscript{13}C]-Valine exhibit an initial increase in enhancement followed by a gradual decline at longer bubbling durations, demonstrating the dependence of signal enhancement on bubbling time under these experimental conditions. Error bars represent the standard deviation of repeated measurements. The number of replicates for each data point is provided in Table \ref{table_S3} (ESI)}
\label{SE_vs_btime_500ac}
\end{figure}

For a different concentration ratio ([Val]/[Cat] = 6:1, Figure\ref{SE_vs_btime_500ac} and Tables \ref{table_S3}, \ref{table_S4}, \ref{table_S5} (ESI), the highest enhancement was observed at the 40 s of bubbling time, followed by a gradual decrease at longer bubbling durations. These results suggest that the optimal bubbling time depends sensitively on experimental conditions, including the substrate-to-catalyst ratio. 

This variability likely reflects the interplay between polarization build-up, relaxation, and catalyst stability under different experimental conditions.

\begin{figure} [!ht]
\includegraphics[width=0.45\textwidth]{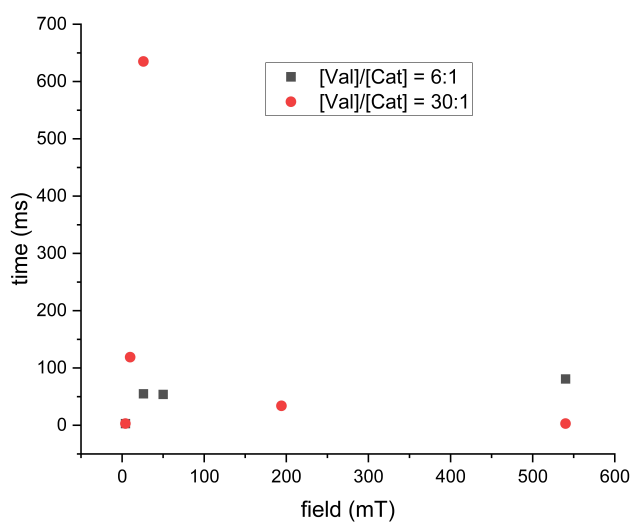}
\caption{Longitudinal relaxation times ($T_1$) of the free L-[1-$^{13}$C]valine carboxyl resonance (approximately 175 ppm) measured using a field-cycling shuttle system on a 400 MHz NMR spectrometer. Relaxation occurred at controlled low magnetic fields, with detection at high field.}
\label{T1}
\end{figure}

To assess relaxation under the corresponding sample conditions, $^{13}$C $T_1$ measurements of the free L-[1-$^{13}$C]valine carboxyl resonance (approximately 175 ppm) were performed on samples recovered from the SABRE hyperpolarization experiments at substrate-to-catalyst ratios of [Val]/[Cat] = 6:1 and 30:1 using an inversion recovery pulse sequence on a 400 MHz NMR spectrometer equipped with a field-cycling shuttle system, allowing relaxation to occur at controlled low magnetic fields while detection was performed at high field, following the general methodology described by Bengs \textit{et al.}\cite{bengs2021nuclear}. As shown in Figure~\ref{T1}, the measured $T_1$ values are short, typically below 1 s under these conditions. These relaxation times are consistent with previously reported low-field $T_1$ values for amino acids\cite{taglang2018late}.

The decrease in signal enhancement at prolonged bubbling times is primarily attributed to catalyst deactivation during continuous exposure to parahydrogen, with rapid relaxation further limiting the observable signal.

The lower signal enhancements observed in the more aqueous solvent are consistent with changes in catalyst speciation and substrate exchange that have previously been reported for parahydrogen-based hyperpolarization systems\cite{bondar2022effect,gater2025perfluorosolvents,min2025waterrecyclablecatalysts}

\subsection{Effect of Solvent Composition on Polarization Efficiency}

\begin{figure} [!ht]
\includegraphics[width=0.45\textwidth]{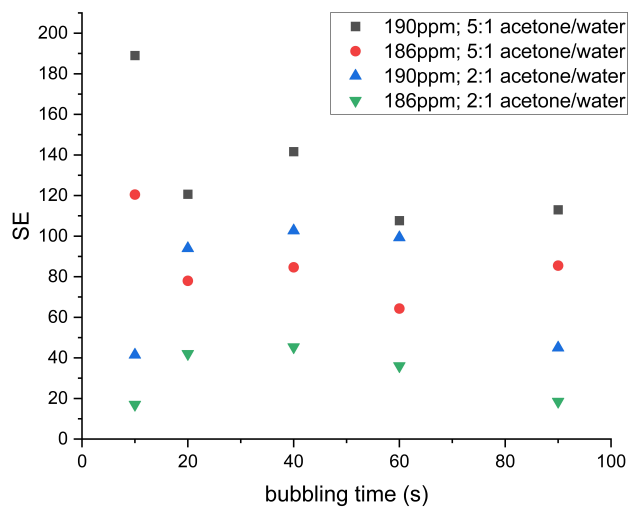}
\caption{Effect of solvent composition on signal enhancement (SE) as a function of bubbling time for L-[1-\textsuperscript{13}C]valine at a substrate-to-catalyst ratio of [Val]/[Cat] = 6:1 ([Val] = 24 mM). Comparison of 5:1 (black squares and red circles) and 2:1 (blue triangles and green inverted triangles) acetone-d6:D\textsubscript{2}O mixtures demonstrates similar time profiles but overall lower signal enhancements in the more aqueous solvent.}
\label{SE_vs_ac}
\end{figure}

To evaluate the influence of solvent composition on SABRE hyperpolarization, the \textsuperscript{13}C signal enhancement of L-[1-\textsuperscript{13}C]-valine was measured in acetone:D$_2$O mixtures with volume ratios of 5:1 and 2:1 while maintaining identical experimental conditions. The results are shown in Figure~\ref{SE_vs_ac}.

For both solvent mixtures, the signal enhancement reached a maximum after approximately 20 s of parahydrogen bubbling and decreased at longer bubbling times. Higher signal enhancements were consistently obtained in the 5:1 acetone:D$_2$O mixture than in the 2:1 mixture. Similar solvent-dependent behavior has been reported previously for parahydrogen-based hyperpolarization systems\cite{bondar2022effect}, and agrees with our previous observations for hyperpolarized pyruvate derivatives\cite{bondar2024hyperpolarised}.

The lower signal enhancement observed in the more aqueous solvent is consistent with previous reports showing that increasing water content can influence catalyst behavior and reduce hyperpolarization efficiency in parahydrogen-based systems\cite{gater2025perfluorosolvents,min2025waterrecyclablecatalysts,bondar2022effect}.

Despite the lower enhancement, detectable SABRE hyperpolarization was achieved in the 2:1 acetone:D$_2$O mixture, demonstrating that polarization transfer remains possible in partially aqueous media.

\subsection{Dependence of Signal Enhancement on Substrate–Catalyst Ratio.}

\begin{figure} [!ht]
\includegraphics[width=0.45\textwidth]{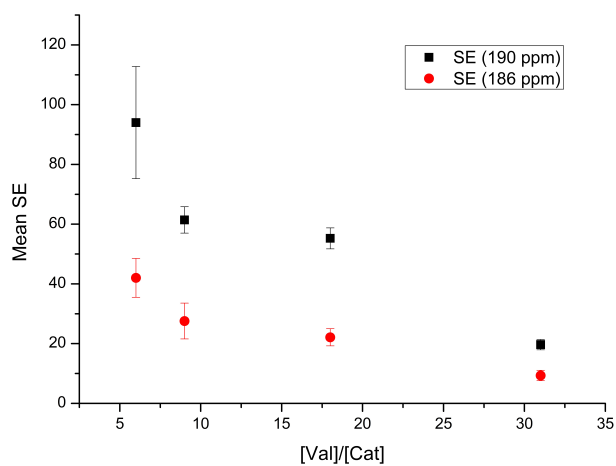}
\caption{Dependence of \textsuperscript{13}C SABRE signal enhancement (SE) on the substrate-to-catalyst concentration ratio. Measurements were performed in acetone:D\textsubscript{2}O (2:1 v/v) with a fixed bubbling time of 20 s.}
\label{SE_vs_conc}
\end{figure}

The relative concentrations of substrate and catalyst also influence SABRE performance. To explore this effect, \textsuperscript{13}C signal enhancements were measured across a range of [Val]/[Cat] ratios while maintaining a constant solvent composition (acetone:D\textsubscript{2}O = 2:1) and fixed bubbling time of 20 seconds (Figure~\ref{SE_vs_conc}).

The apparent decrease in signal enhancement observed at higher substrate-to-catalyst ratios should be interpreted with caution. Because the hyperpolarized resonances originate from catalyst-associated valine species whereas the enhancement is referenced to the thermal signal of free L-[1-\textsuperscript{13}C]valine, increasing the substrate concentration may increase the thermal reference signal without a proportional increase in the population of catalyst-bound species. Saturation of the available catalyst binding sites could therefore contribute to the observed decrease in the apparent signal enhancement. Further investigation over a broader range of substrate-to-catalyst ratios would be required to establish the underlying mechanism.

The increase in SE at lower ratios likely reflects improved substrate exchange at the active catalyst sites and more efficient polarization transfer as the concentration of valine increases. However, beyond the optimal point, excess valine may begin to compete with parahydrogen for binding sites on the catalyst or disrupt the optimal coordination geometry required for polarization transfer. High substrate concentrations can also lead to catalyst aggregation or changes in chemical exchange dynamics, reducing the efficiency of spin transfer.

These results demonstrate that the substrate-to-catalyst ratio is an important experimental parameter influencing the observed signal enhancement.

\subsection{Preliminary SABRE Hyperpolarization of L-[1-\textsuperscript{13}C]Glycine}

\begin{figure}[!ht]
\centering
\includegraphics[width=0.95\linewidth]{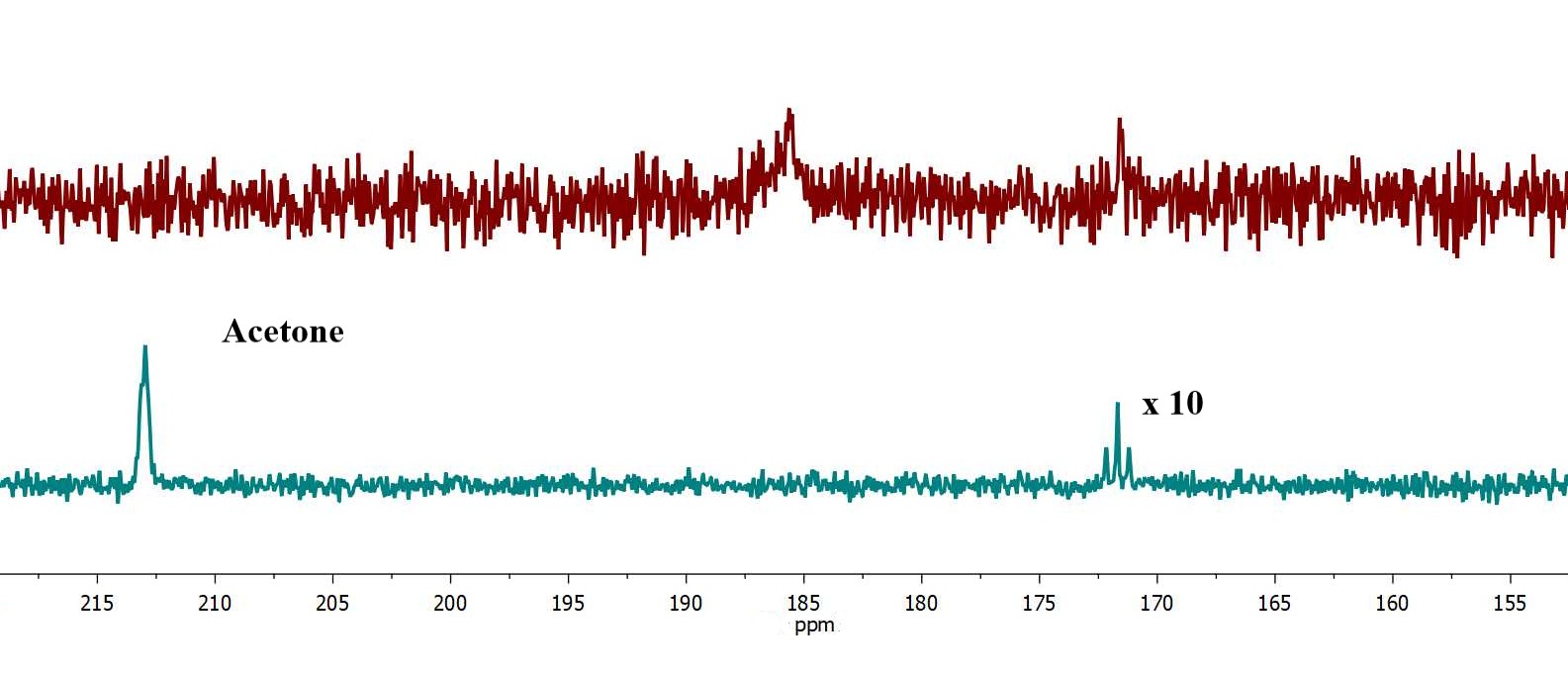}
\caption{Representative thermal (bottom) and SABRE-hyperpolarized (top) \textsuperscript{13}C NMR spectra of L-[1-\textsuperscript{13}C]glycine acquired at 1.1~T. Hyperpolarized resonances are observed at approximately 185 and 171~ppm. The resonance at 171~ppm coincides with the thermal resonance of free L-[1-\textsuperscript{13}C]glycine.}
\label{fig:Gly_spectra}
\end{figure}

To evaluate whether the SABRE methodology developed for L-[1-\textsuperscript{13}C]valine could be extended to another unmodified amino acid, preliminary experiments were performed using L-[1-\textsuperscript{13}C]glycine under comparable experimental conditions. Representative thermal and hyperpolarized \textsuperscript{13}C NMR spectra are shown in Figure~\ref{fig:Gly_spectra}.

Two hyperpolarized resonances were observed at approximately 185 and 171~ppm. The resonance at 171~ppm coincides with the chemical shift of free L-[1-\textsuperscript{13}C]glycine in the thermal spectrum, whereas the resonance at approximately 185~ppm is tentatively assigned to a catalyst-associated species. In contrast to L-[1-\textsuperscript{13}C]valine, glycine exhibited hyperpolarization of both a catalyst-associated species and the free amino acid resonance, indicating differences in substrate coordination and exchange under otherwise comparable SABRE conditions. Although glycine-containing probes have previously been investigated using dissolution dynamic nuclear polarization (dDNP) for metabolic imaging through enzymatic generation of hyperpolarized [1-\textsuperscript{13}C]glycine\cite{batsios2020vivo}, the present study demonstrates direct SABRE hyperpolarization of unmodified L-[1-\textsuperscript{13}C]glycine without chemical derivatization. These complementary approaches highlight the potential of glycine as a biologically relevant hyperpolarized probe while relying on fundamentally different polarization strategies.

\begin{figure}[!ht]
\centering
\includegraphics[width=0.95\linewidth]{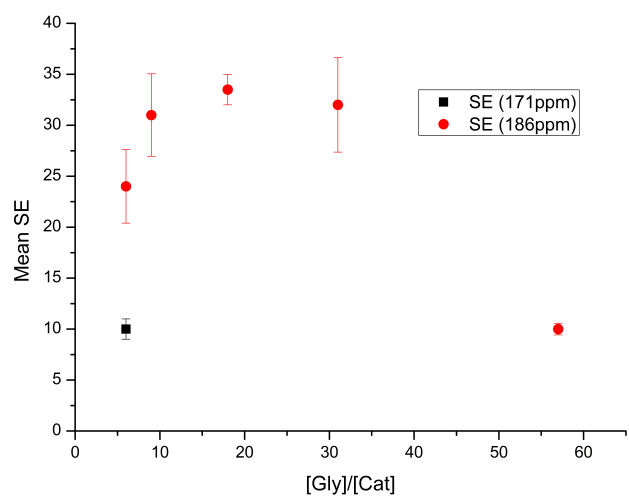}
\caption{Dependence of the \textsuperscript{13}C signal enhancement on the substrate-to-catalyst ratio for L-[1-\textsuperscript{13}C]glycine. Error bars represent the standard deviation of repeated measurements.}
\label{fig:Gly_SE_vs_conc}
\end{figure}

The influence of the substrate-to-catalyst ratio on signal enhancement at a fixed parahydrogen bubbling time of 20~s is shown in Figure~\ref{fig:Gly_SE_vs_conc}. The resonance at approximately 185~ppm was observed over the investigated concentration range, whereas the resonance at approximately 171~ppm was detected only at the lowest substrate-to-catalyst ratio investigated ([Gly]/[Cat] = 6:1). These preliminary observations suggest that the substrate-to-catalyst ratio influences the formation and exchange of the hyperpolarized species.

\section{Conclusions}
We demonstrate SABRE-mediated \textsuperscript{13}C hyperpolarization of unmodified L-[1-\textsuperscript{13}C]-valine in a partially aqueous solvent system at room temperature. Signal enhancements of up to approximately 60-fold were observed using a benchtop 1.1 T NMR spectrometer, confirming the feasibility of applying SABRE to zwitterionic amino acids without chemical modification.

The presence of multiple hyperpolarized resonances suggests that polarization is associated with catalyst-bound species, highlighting the importance of substrate–catalyst interactions in determining polarization transfer pathways.

The dependence of signal enhancement on bubbling time, solvent composition, and substrate-to-catalyst ratio demonstrates that SABRE performance in such systems is highly sensitive to experimental conditions. Rapid relaxation and catalyst deactivation are likely to contribute to the observed decrease in signal enhancement. The observed dependence of signal enhancement on substrate-to-catalyst ratio highlights the importance of catalyst optimization in parahydrogen-based hyperpolarization experiments, in agreement with previous studies on hydrogenative PHIP systems\cite{di2025improving}.

Overall, this work provides a proof-of-concept for extending SABRE hyperpolarization to unmodified amino acids in partially aqueous media and establishes a foundation for future investigations in more biologically relevant environments. 

\section{Materials and Methods}

All reagents were obtained from commercial suppliers and used without further purification unless otherwise stated. L-[1-\textsuperscript{13}C]valine and L-[1-\textsuperscript{13}C]glycine ($\geq$99 atom \% \textsuperscript{13}C) together with the solvents (acetone-d\textsubscript{6}, D\textsubscript{2}O and methanol-d\textsubscript{4}, as applicable) were purchased from commercial suppliers. The SABRE precatalyst, [Ir(IMes)(COD)Cl] (where IMes = 1,3-bis(2,4,6-trimethylphenyl)imidazol-2-ylidene; COD = 1,5-cyclooctadiene), was synthesized following literature protocols.

Sample preparation was performed under ambient conditions. All solutions were prepared in 5 mm Young's capped NMR tubes (5 mm Precision NMR Sample Tube, Low Pressure/Vacuum Valve (LPV) 8L, 500 MHz rated). The final sample volume was typically 600 $\mu$L. Unless otherwise stated, experiments were performed using L-[1-\textsuperscript{13}C]valine dissolved in acetone-d\textsubscript{6}/D\textsubscript{2}O mixtures together with the iridium catalyst. Substrate concentration, substrate-to-catalyst ratio and solvent composition were varied as described for the individual experiments. Preliminary glycine experiments were performed under comparable experimental conditions.

Parahydrogen (92\% enrichment, Bruker Parahydrogen Generator) was bubbled through the solution for 10–60 seconds, as indicated in each experiment. Bubbling was performed inside a $\mu$-metal shielded solenoid coil (internal B\textsubscript{0} $\approx$ 9 mG), at room temperature, to maintain optimal polarization transfer conditions.

After bubbling, the samples were rapidly transferred (within $\approx$ 2 seconds) to a 1.1 T benchtop NMR spectrometer (Magritek Spinsolve 43 MHz) for immediate acquisition. Hyperpolarized spectra were recorded using a single scan, while thermal spectra were acquired using signal averaging over 256–1024 scans with a recycle delay long enough for 5 times T\textsubscript{1}, depending on the signal-to-noise ratio.

Further experimental details, data analysis procedures, and supporting figures are provided in the ESI.

\section*{Author Contributions}
O. B.: Conceptualization, Formal analysis, Investigation, Methodology, Visualization, Writing – original draft, Writing – review \& editing. T. B. R. R.: Supervision, Writing – review \& editing.

\section*{Conflicts of interest}
 ``There are no conflicts to declare''.

\section*{Acknowledgements}
We thank the British Academy for funding support and the School of Chemistry of the University of Southampton for supporting Dr. Bondar by providing infrastructure support.

\section*{Data availability}
The processed NMR spectra, individual signal enhancement measurements, and numerical data used for the statistical analyses are provided within the article and the Electronic Supplementary Information. The original raw NMR acquisition files are archived on the University of Southampton research systems and are not publicly available.



\balance


\bibliography{rsc} 
\bibliographystyle{rsc} 

\clearpage
\section*{Electronic Supplementary Information}
\input{SI.tex}

\end{document}

%% file: SI.tex



\setcitestyle{super}
\title{Electronic Supplementary Information (ESI): SABRE Hyperpolarization of Unmodified Amino Acids in Partially Aqueous Media: L-[1-\textsuperscript{13}C]-Valine as a Model System.}

\footnotetext{\textit{$^{a}$~School of Chemistry, Highfield Campus, Southampton, SO17 1BJ United Kingdom; E-mail: oksana.BONDAR@soton.ac.uk}}
\footnotetext{\textit{$^{b}$~Address, Address, Town, Country. }}

\section*{SABRE Hyperpolarization Setup Diagram}
\begin{figure} [H]
    \centering
    \includegraphics[width=1\linewidth]{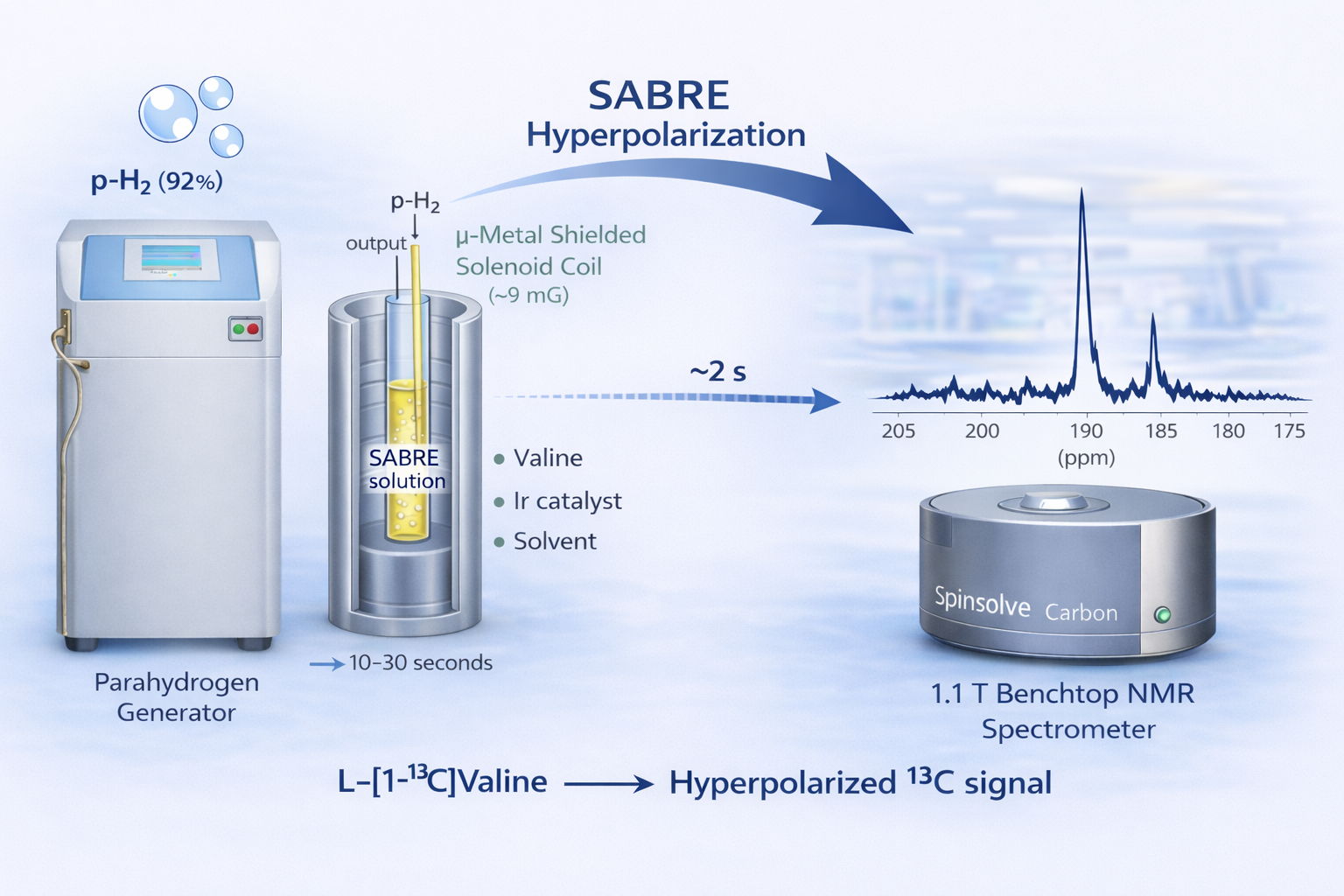}
    \caption{Schematic representation of the SABRE hyperpolarization setup. Parahydrogen (p-H\textsubscript{2}) generated using a Bruker parahydrogen generator is bubbled through a solution containing L-[1-\textsuperscript{13}C]valine, the iridium catalyst, and solvent inside a $\mu$-metal shielded solenoid coil ($\sim$ 9 mG). After polarization transfer (10–30 s bubbling), the sample is rapidly transferred ($\sim$2 s) to a 1.1 T Spinsolve Carbon benchtop NMR spectrometer for detection of the hyperpolarized \textsuperscript{13}C signal.}
    \label{fig:placeholder}
\end{figure}

\section*{Materials and Methods}

Signal enhancement (SE) was calculated as the ratio of the hyperpolarized and thermal signal integrals,

\begin{equation}
SE=\frac{S_{\mathrm{hyp}}}{S_{\mathrm{therm}}}
\end{equation}

where $S_{\mathrm{hyp}}$ is the integral of the hyperpolarized resonance and $S_{\mathrm{therm}}$ is the integral of the corresponding thermal resonance when observable.

When no corresponding thermal resonance was detected at the same chemical shift, the detectable thermal resonance of the corresponding free amino acid was used as the reference. For L-[1-\textsuperscript{13}C]valine, the thermal resonance at approximately 175 ppm was used because the hyperpolarized resonances at approximately 190 and 186 ppm had no directly observable thermal counterparts. The same approach was used for the preliminary L-[1-\textsuperscript{13}C]glycine experiments using the thermal resonance of free L-[1-\textsuperscript{13}C]glycine where appropriate.

Because the hyperpolarized resonances at approximately 190 and 186 ppm do not have directly observable thermal counterparts, the resulting signal enhancement values should be regarded as approximate estimates for comparing relative SABRE performance under different experimental conditions and should not be interpreted as direct measurements of the polarization gain of the individual hyperpolarized species.

Spectra were processed using Spinsolve Expert (v2.01.08), and all signal integrals were automatically normalized for the number of scans using the software's built-in processing tools.

Longitudinal relaxation times ($T_1$) were measured for the valine samples using an inversion recovery pulse sequence with \textsuperscript{13}C detection.

\section*{Effect of Parahydrogen Bubbling Time on Signal Enhancement}
Individual signal enhancement measurements used for calculation of the mean values and standard deviations reported in the main text are listed below. 

\begin{table*} [ht]
\small
    \centering
    \caption{Individual signal enhancement measurements used to calculate the values reported in Table 1 at a main text ($\sim$190 ppm) of L-[1-\textsuperscript{13}C]-Valine}
\label{table_S1}
    \begin{tabular}{c|c|c|c|c|c|c|c}
 \hline\
  Bubbling time & SE(Replicate 1) & SE(Replicate 2) &  SE(Replicate 3) & SE(Replicate 4) & SE(Replicate 5) & SE(Replicate 6) & SE(Replicate7)\\  
 \hline\hline
 10	& 25 & 17\\
\hline\
20	& 19 & 39 & 24 & 24 & 21 & 48 & 18\\
\hline
40	& 29 & 30 & 18 & 27 & 24 & 29\\
\hline
60 & 21 & 16 & 26 & 19\\
\hline
90 & 19 & 27 & 25 & 19 \\
\hline
\end{tabular}
\end{table*}

\begin{table*} [ht]
\small
    \centering
    \caption{Individual signal enhancement measurements used to calculate the values reported in Table 3 at a main text ($\sim$186 ppm) of L-[1-\textsuperscript{13}C]-Valine}
\label{table_S2}
    \begin{tabular}{c|c|c|c|c|c|c|c}
 \hline\
  Bubbling time & SE(Replicate 1) & SE(Replicate 2) &  SE(Replicate 3) & SE(Replicate 4) & SE(Replicate 5) & SE(Replicate 6) & SE(Replicate7)\\  
 \hline\hline
 10	& 57 & 40\\
\hline\
20	& 46 & 61 & 56 & 52 & 58 & 112 & 45\\
\hline
40	& 58 & 55 & 40 & 64 & 54 & 52\\
\hline
60 & 48  & 37 & 53 & 48 \\
\hline
90 & 43 & 38 & 66 & 47 \\
\hline
\end{tabular}
\end{table*}

To further investigate the influence of bubbling duration on SABRE polarization efficiency, additional experiments were performed at a substrate-to-catalyst ratio of [Val]/[Cat] = 6:1 over an extended bubbling time range. These measurements allow evaluation of the polarization build-up and decay behavior beyond the time window explored in the main manuscript. The results confirm that signal enhancement initially increases with bubbling time before decreasing at longer durations due to relaxation and catalyst dynamics.

\begin{table*} [ht]
\small
\centering
\caption{Summary of signal enhancement (SE) values as a function of parahydrogen bubbling time for the two observed \textsuperscript{13}C resonances ($\sim$190 ppm and $\sim$186 ppm) of L-[1-\textsuperscript{13}C]-Valine at a substrate-to-catalyst ratio of [Val]/[Cat] = 6:1 (Figure4 main text).}
\label{table_S3}
    \begin{tabular}{c|c|c|c|c|c}
 \hline\
  Bubbling time & SE(bound Valine at 190 ppm) & Error SE at 190 ppm &  SE(bound Valine at 186 ppm) & Error SE at 186 ppm & n(repeats) \\  
 \hline\hline
 10	& 41.5 & 1.5	& 17 & 4 & 2 \\
\hline\
20	& 94 & 18.7705443 & 42	& 6.506407099 & 3 \\
\hline
40	& 102.6666667 & 13.29578045	& 45.33333333 & 5.364492313	& 3 \\ 
\hline
60 & 99.33333333 & 13.73964256 & 36	& 0.5	& 3 \\
\hline
90 & 45	& 5.567764363 & 18.5 & 0.5 & 3 \\
\hline
\end{tabular}
\end{table*}

\begin{table*} [ht]
\small
    \centering
    \caption{Individual signal enhancement measurements used to calculate the values reported in Table 1 at a main text ($\sim$190 ppm) of L-[1-\textsuperscript{13}C]-Valine}
\label{table_S4}
    \begin{tabular}{c|c|c|c}
 \hline\
  Bubbling time & SE(Replicate 1) & SE(Replicate 2) &  SE(Replicate 3)\\  
 \hline\hline
 10	& 43 & 40\\
\hline\
20	& 126 & 61 & 95\\
\hline
40	& 83 & 128 & 97\\
\hline
60 & 78 & 95 & 125\\
\hline
90 & 34 & 49 & 52 \\
\hline
\end{tabular}
\end{table*}

\begin{table*} [ht]
\small
    \centering
    \caption{Individual signal enhancement measurements used to calculate the values reported in Table 1 at a main text ($\sim$186 ppm) of L-[1-\textsuperscript{13}C]-Valine}
\label{table_S5}
    \begin{tabular}{c|c|c|c}
 \hline\
  Bubbling time & SE(Replicate 1) & SE(Replicate 2) &  SE(Replicate 3)\\  
 \hline\hline
 10	& 21 & 13\\
\hline\
20	& 49 & 29 & 48\\
\hline
40	& 39 & 56 & 41\\
\hline
60 & 30 & 29 & 49 \\
\hline
90 &  & 18 & 19 \\
\hline
\end{tabular}
\end{table*}

\section*{Preliminary SABRE Hyperpolarization of L-[1-\textsuperscript{13}C]Glycine}

To investigate whether the SABRE methodology developed for L-[1-\textsuperscript{13}C]valine could be extended to another unmodified amino acid, preliminary experiments were performed using L-[1-\textsuperscript{13}C]glycine under experimental conditions comparable to those employed throughout this study. These experiments were intended as an initial assessment of the applicability of the method and were therefore not subjected to the same level of optimization as the valine measurements presented in the main manuscript.

\begin{figure}[!ht]
    \centering
    \includegraphics[width=0.75\linewidth]{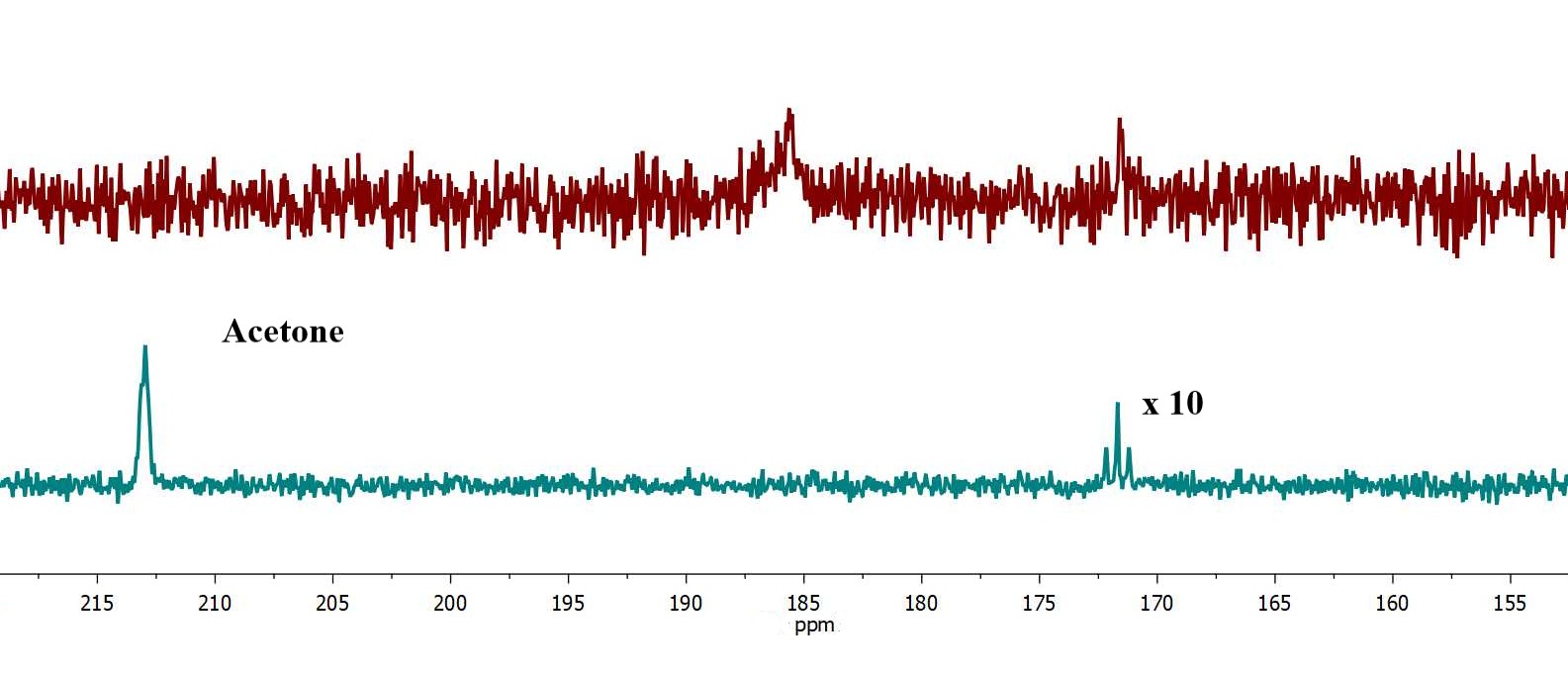}
    \caption{
Representative thermal (bottom) and SABRE-hyperpolarized (top) \textsuperscript{13}C NMR spectra of L-[1-\textsuperscript{13}C]glycine acquired under the experimental conditions described in the main manuscript. Hyperpolarized resonances are observed at approximately 185 and 171~ppm. The resonance at 171~ppm coincides with the thermal resonance of free L-[1-\textsuperscript{13}C]glycine.
}
    \label{fig:Gly_spectra}
\end{figure}

Representative thermal and hyperpolarized \textsuperscript{13}C NMR spectra are shown in Figure~\ref{fig:Gly_spectra}. Two hyperpolarized resonances were observed at approximately 185 and 171~ppm. The resonance at 171~ppm coincides with the chemical shift of free L-[1-\textsuperscript{13}C]glycine observed in the thermal spectrum, whereas the resonance at approximately 185~ppm is tentatively assigned to a catalyst-associated species. These observations indicate that glycine exhibits SABRE behaviour distinct from that of L-[1-\textsuperscript{13}C]valine under otherwise comparable experimental conditions.

\begin{figure}[!ht]
    \centering
    \includegraphics[width=0.75\linewidth]{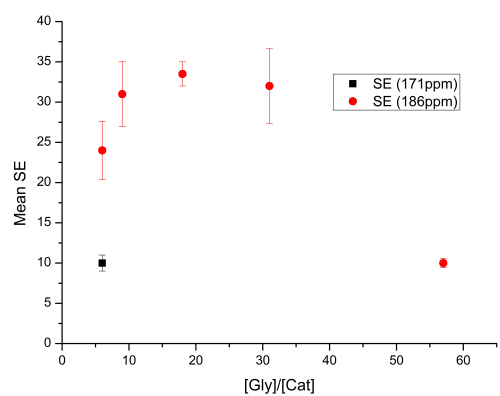}
    \caption{
Signal enhancement as a function of the substrate-to-catalyst ratio for L-[1-\textsuperscript{13}C]glycine measured at a fixed parahydrogen bubbling time of 20~s. Error bars represent the standard deviation of repeated measurements.
}
    \label{fig:Gly_SE_vs_conc}
\end{figure}

The influence of the substrate-to-catalyst ratio on the signal enhancement at a fixed parahydrogen bubbling time of 20~s is shown in Figure~\ref{fig:Gly_SE_vs_conc}. The resonance at approximately 185~ppm was observed over the investigated concentration range, whereas the resonance at approximately 171~ppm was detected only at the lowest substrate-to-catalyst ratio investigated ([Gly]/[Cat] = 6:1). These preliminary results suggest that the coordination and exchange behaviour of glycine differs from that of valine under otherwise comparable SABRE conditions.

\begin{figure}[!ht]
    \centering
    \includegraphics[width=0.95\linewidth]{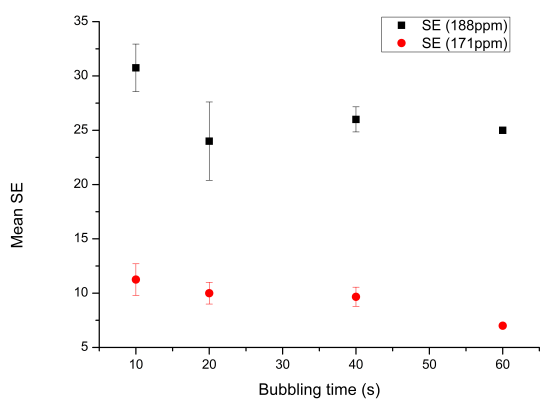}
    \caption{
Signal enhancement as a function of parahydrogen bubbling time for L-[1-\textsuperscript{13}C]glycine measured at a fixed substrate-to-catalyst ratio of 6:1. Error bars represent the standard deviation of repeated measurements where applicable.
}
    \label{fig:Gly_SE_vs_b.time}
\end{figure}

The influence of parahydrogen bubbling time on the signal enhancement of L-[1-\textsuperscript{13}C]glycine at a fixed substrate-to-catalyst ratio of 6:1 is shown in Figure~\ref{fig:Gly_SE_vs_b.time}. As observed for L-[1-\textsuperscript{13}C]valine, the signal enhancement depends on bubbling duration, indicating that polarization transfer is influenced by the balance between hydrogenation, ligand exchange, and spin relaxation processes.

\begin{figure}[!ht]
    \centering
    \includegraphics[width=0.75\linewidth]{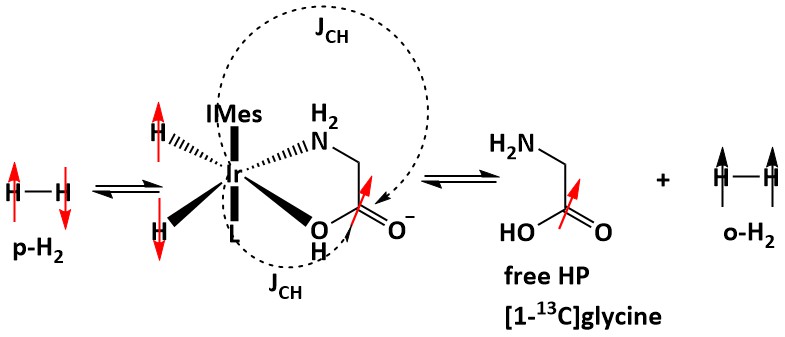}
    \caption{
Proposed reversible coordination pathway for L-[1-\textsuperscript{13}C]glycine during SABRE hyperpolarization. The scheme is presented as a qualitative model consistent with the experimental observations and is not intended as a definitive structural assignment.
}
    \label{SABRE_Gly}
\end{figure}

A schematic representation of a possible reversible coordination pathway consistent with the observed hyperpolarized resonances is shown in Figure~\ref{SABRE_Gly}. The proposed coordination model is presented solely as a qualitative illustration of a plausible SABRE exchange process consistent with the experimental observations and should not be regarded as a definitive structural assignment of the observed species.

